\documentclass[11pt,reprint]{revtex4-1}

\usepackage[usenames,dvipsnames]{color}            
\newcommand{\wf}[1]{}
\newcommand{\wk}[1]{}

\usepackage{graphicx,epsfig,amssymb,amsbsy,amsmath,amsthm,cleveref,enumitem}
\newcommand{\bb}[1]{\left({#1}\right)}					
\newcommand{\sq}[1]{\left[#1\right]}						
\newcommand{\cc}[1]{\left\{#1\right\}}					
\newcommand{\op}[1]{\mathcal{#1}}
\newcommand{\ord}[1]{{\sf O}\mbox{\small$\bb{#1}$}}					
		
\newcommand{\sfrac}[2]{\mbox{$\frac{#1}{#2}$}}	
\newcommand{\hf}{\mbox{$\sfrac12$}}

\renewcommand{\v}[1]{{\bf #1}}							

\newcommand{\uuline}[1]{\underline{\underline{#1}}}

\newcommand{\im}{{\operatorfont i}}									
\newcommand{\id}{1}				

\newcommand{\eps}{\varepsilon}

\newcommand{\mirror}[1]{\overline{#1}}

\newcommand{\adj}[1]{{{#1}^\dagger}}

\newcommand{\reals}{\mathbb{R}}						
\newcommand{\imags}{\mathbb{I}}						
\newcommand{\rep}[1]{\reals\sq{#1}}					
\newcommand{\imp}[1]{\imags\sq{#1}}					

\newcommand{\e}[1]{{\mbox{\boldmath$e$}}_{#1}}						
\newcommand{\ec}[1]{{\mbox{\boldmath$e$}}^{#1}}						
\newcommand{\tayl}{Y}								
\newcommand{\con}{\Gamma}
\newcommand{\tens}{\uuline}

\renewcommand{\tayl}[2]{{\sf tayl}}

\newcommand{\m}{M}
\newcommand{\n}{N}

\newtheorem{example}{Example}

\crefname{equation}{}{}
\Crefname{equation}{}{}

\def\eps{\varepsilon}

\begin{document}

\title{Higher space-time symmetries from simple mixture}
\title{Symmetries from mixtures: an algebra for gauge theories}
\author{Mike R. Jeffrey}
\address{Engineering Mathematics, University of Bristol, Merchant Venturer's Building, Bristol BS8 1UB, UK, email: mike.jeffrey@bristol.ac.uk}
\date{\today}

\begin{abstract}\noindent
We present here a product between vectors and scalars that {\it mixes} them within their own space, using imaginaries to describe geometric products between vectors, forming complex vectors rather than introducing higher order/dimensional vector objects. This is done by means of a {\it mixture tensor} that captures the rich geometries of geometric algebras, and simultaneously lends itself naturally to tensor calculus. We use this to develop a notion of analyticity in higher dimensions based on the idea that a function can be made differentiable --- in a certain strong sense --- by permitting curvature of the underlying space, and we call this {\it analytic curvature}. 

To explore these ideas we use them to derive a few fundamental laws and operators of physics which, while considered somewhat lightly, have compelling features. The mixture, for instance, produces rich symmetries without adding dimensions beyond the familiar space-time, and its derivative produces familiar quantum field relations in which the field potentials are just derivatives of the coordinate basis. The symmetries of physical laws are seen to arise not directly from the bases, but from their mixtures. 
\end{abstract}

\maketitle



The are a number of symmetries and group structures known to underly the various laws of fundamental physics, in so far as we understand them. Where those structures come from, how complete they are, and how they are related across the standard model, quantum field theories, and gravitation, for example, remain incompletely understood. A different approach to the algebra and geometry used to describe them may yield a more complete picture. 

Here we approach differential geometry a little differently to known standards. We begin with an algebra based neither around a `scalar' nor `exterior' product, but instead built on a {\it mixture} product that multiplies vectors within their own space. We investigate the differentiability of functions within the algebra so created, and illustrate the formalism by using it to obtain some familiar differential relations of fundamental physical fields. Though proposed rather informally, these reveal rich symmetries and perturbations of potential interest in physical law. These ideas revisit various ideas that have been proposed over the last century, but framed somewhat differently by the algebra we introduce. More detailed calculations and more in-depth investigation of these ideas are set out in \cite{rasp2020long}. 

We start by defining the {\it mixture} product between a general pair of vector bases $\e\beta$ and $\e\gamma$ by means of a tensor $\eta$, as a vector $\e\beta\e\gamma=\eta_\alpha(\e\beta,\e\gamma)$, so the product is a new vector $\e\alpha$. 
Given a system of $n$ orthogonal basis vectors $\cc{\e\alpha}_{\alpha=0,1,2,...,n}\;$, some dual basis $\cc{\ec\alpha}_{\alpha=0,1,2,...,n}\;$, and a {mixture tensor} $\eta$ whose components are real or complex scalars, let the products of bases be given by 
\begin{equation}\label{eta}
\e\alpha \e\beta=\eta^\gamma_{\alpha\beta}\e\gamma\qquad\mbox{and}\qquad
\ec\alpha \ec\beta=\eta_\gamma^{\alpha\beta}\ec\gamma\;. 
\end{equation}
With this we can just as easily define products between vectors and duals via 
$\e\beta\ec\alpha=\eta_{\gamma\beta}^\alpha\ec\gamma=\eta_\beta^{\alpha\gamma}\e\gamma$ and 
$\ec\alpha\e\beta=\eta_{\beta\gamma}^\alpha\ec\gamma=\eta_\beta^{\gamma\alpha}\e\gamma$. 
(We use the convention of summing over repeated upper-lower index pairs, i.e. summing over $\gamma=0,1,...,n$ in these expressions). We might describe this as a `functional' rather than `geometric' algebra, perhaps, as it facilitates the description of algebraic functions of vector quantities, and their calculus. 

Scalar and vector products arise naturally, bearing some similarity to quaternion algebras or Clifford algebras, but with a particular difference, namely that real or imaginary values of the components $\eta_{\alpha\beta}^\gamma$ distinguish between products giving scalar (magnifying) quantities (real) or rotations (imaginary). 

The mixture combines concepts of commutative and non-commutative algebras in one. Its symmetric part is just related to the metric as
\begin{equation}\label{glow}
g_{\alpha\beta}
=\hf(\e\alpha\;\mirror{\e\beta}+\e\beta\;\mirror{\e\alpha})
=\hf(\eta_{\alpha\mirror\beta}^\gamma+\eta_{\beta\mirror\alpha}^\gamma)\e\gamma\;,
\end{equation}
where $\mirror{\e\alpha}=-\e\alpha$ if $\e\alpha$ is non-commutative. 
Antisymmetric products produce mixture components $\eta_{\alpha\beta}^\gamma$ that are antisymmetric in their indices, similar to `structure coefficients' or Levi-Civita symbols of non-commutative bases. 
By combining the metric and antisymmetric tensors, the mixture is only a small augmentation of standard concepts, and indeed such a combined product may well have been developed elsewhere that the author is not aware of. Its real interest, however, is revealed when we consider its effect on calculus. 

The mixture permits us to re-consider tensor calculus for functions and variables, and their derivatives, that all belong to one and the same space. It turns out (see \cite{rasp2020long}) that such derivatives always carry an amount of non-uniqueness (for all but real or complex scalar spaces), but uniqueness can be restored by permitting curvature of the underlying space. 

To this end we permit variation of the bases $\e\gamma$ across the space. For a function $f$ define a derivative with respect to a variable $z$ by
\begin{align}
\sfrac{df}{dz}=\lim_{\Delta z\rightarrow0}\Delta z^{-1}\bb{f(z+\Delta z)-f(z)}\;,
\end{align}
(a corresponding theory can be based around the `right' derivative using $\bb{f(z+\Delta z)-f(z)}\Delta z^{-1}$). 
Here the multiplication is performed via the mixture, while differentiation is performed covariantly. 
We can then write
 $\sfrac{df}{dz}=(\sfrac{df}{dz})^\gamma\e\gamma$, where
\begin{align}\label{dfcon}
(\sfrac{df}{dz})^\gamma=
\eta^{\gamma\beta}_\alpha f^\alpha_{;\beta}
&=\eta^{\gamma\beta}_\alpha( \partial_{\beta}f^\alpha+ \con^\alpha_{\lambda\beta} f^\lambda)\;,
\end{align}
Defining $H^\gamma_\lambda=\eta^{\gamma\beta}_\alpha \con^\alpha_{\lambda\beta}$, this simplifies to
\begin{align}\label{dfH}
(\sfrac{df}{dz})^\gamma=
\eta^{\gamma\beta}_\alpha f^\alpha_{;\beta}
&=(\eta^{\gamma\beta}_\alpha \partial_{\beta}+ H^\gamma_\alpha)f^\alpha\;.\quad 
\end{align}
We will apply this below to derive the Dirac equations as analyticity conditions. But first, taking \cref{dfcon}, if $\con$ is symmetric this derivative can be written
\begin{align}
(\sfrac{df}{dz})^\gamma=\eta^{\gamma\beta}_\alpha f^\alpha_{;\beta}&=\eta^{\gamma\beta}_\lambda(1\partial_{\beta}+\op G_{\beta})^\lambda_\alpha f^\alpha\;,
\end{align}
where $\op G_\mu$ is a square matrix with components $(\op G_\mu)^\alpha_\beta=\con^\alpha_{\mu\beta}$, and $1$ is the identity matrix. This form is consistent with the derivative of the standard model of particle physics if $\op G_\mu=\epsilon\op H_\mu$ where $\op H_\mu$ are field potentials and $\epsilon$ a coupling parameter. The curvature tensor is then
\begin{align}\label{Ryang}
R^{\alpha}_{\beta\nu\mu}&=2\epsilon(\op H^{}_{[\mu,\nu]}+\epsilon\op H^{}_{[\nu}\op H^{}_{\mu]})^\alpha_\beta
\;:=\;2\epsilon(F_{\mu\nu})^\alpha_\beta\;,
\end{align}
giving the electromagnetic field tensor $F^{\mu\nu}$ of Yang-Mills' theory (see e.g. \cite{cg07,yangmills}). The symmetries and the richness of gauge invariance of physics then arise not from the basis of a space directly, and so do not require adding dimensions beyond the familiar $3+1$ of space-time. Instead the rich symmteries arise from the presence of the mixture in these expressions, along with the covariant derivative. 

A special case is given if the electromagnetic potential $h_\alpha=\cc{\phi,\v A}$ is just the divergence of the coordinate basis, 
\begin{align}
h_\alpha\ec\alpha=\ec\alpha_{,\alpha} \;,
\end{align}
then the potentials are just the trace of the connection, $h_\alpha=\con^\beta_{\alpha\beta}$, 
and the curvature tensor is related to the electromagnetic field (Faraday) tensor $F_{\mu\nu}$ as 
\begin{align}\label{Rpauli}
R^\gamma_{\alpha\mu\nu}=\id^\gamma_\alpha h_{[\mu,\nu]}=\id^\gamma_\alpha F_{\nu\mu}\;,
\end{align}
a result reminiscent of an association of the Riemann tensor $R$ with the Faraday tensor $F$ once proposed by Pauli. 

The results \cref{Ryang} and \cref{Rpauli} themselves do not require the mixture algebra introduced above, but the algebra does alter the way we think about coordinate bases, and variables or functions built on them. 
Imaginary quantities, in particular, play an inescapable role associated with anti-commutative bases and rotations. If $\e\alpha$ and $\e\beta$ anticommute then $(\e\alpha\e\beta)^2$, and if $(\e\gamma)^2=1$ for any bases, the simplest mixture is given by $\e\alpha\e\beta=\im\e\gamma$. If we accept such a role for imaginary quantities, we are led, though somewhat more speculatively, to consider weak fields as perturbations of flat space involving not only real, but imaginary, quantities. 

Let us propose that the metric of flat space is perturbed by a weak gravitational field potential $\psi$, via a real perturbation, and by a weak electromagnetic field potential $\phi\ec0+A_j\ec j$, via an imaginary perturbation, say as 
\begin{align}
g_{00}&=1/g^{00}=-1-\mu_g\psi-\im\mu_e\phi\;,\nonumber\\ 
g^{jj}&=\;\;g_{jj}\;\;=+1+\mu_g\psi+\im\mu_e\phi\;,\\\nonumber
g_{0j}&=\;\;g^{0j}\;=\hf\im\mu A_j\;,
\end{align}
and $g^{jk}=g_{jk}=0$ for $j\neq k$, for some constants $\mu_g,\mu_e$. 
(These can easily be written directly as perturbations of the bases $\e\alpha$, but we will try to remain as close to standard theory as possible). 
The usual formula for the metric applies, namely
\begin{equation}\label{metstand}
\con^\sigma_{\alpha\mu}=\hf g^{\beta\sigma}(g_{\alpha\beta,\mu}+g_{\beta\mu,\alpha}-g_{\mu\alpha,\beta})\;,
\end{equation}
and for the $\mu=0$ component this yields
\begin{align}
\con^\omega_{\gamma0}=\mu_e( J^\omega_\gamma\;+\;\hf\im F^{\omega\lambda}\id_{\lambda\gamma})\;,\qquad\;\;
\end{align}
consisting of a gravitational contribution $J$ and electromagnetic contribution $F$, where
\begin{align}
J^\omega_\gamma&=\mbox{\footnotesize$\bb{\begin{array}{cccc}\;\;0\;\;&\;\;G_1\;\;&\;G_2\;\;&\;\;G_3\hspace{0.04cm}\\G_1&0&0&0\\G_2&0&0&0\\G_3&0&0&0\end{array}}  $}\;,\\
F^{\omega\lambda}\id_{\lambda\gamma}&=\mbox{\footnotesize$\bb{\begin{array}{cccc}0&E_1&E_1&E_3\\-E_1&0&-B_3&B_2\\-E_2&B_3&0&-B_1\\-E_3&-B_2&B_1&0\end{array}}  $}\;,
\end{align}
with $G=\nabla\psi$, $E=\nabla\phi+\partial_t A$, $B=\nabla\times A$, in terms of the gravitational potential $\psi$ and electromagnetic 4-potential $(\phi,A)$.

We can be much more general. If $\con$ is not symmetric on its lower indices then in place of \cref{metstand} we have  
\begin{equation}
\con^\sigma_{\alpha\mu}=\mbox{``\cref{metstand}''}
+\hf g^{\beta\sigma}(C_{\beta\alpha\mu}+C_{\beta\mu\alpha}-C_{\alpha\mu\beta})\;
\end{equation}
where $C_{\beta\alpha}^\delta=g^{\mu\delta}C_{\beta\alpha\mu}$ are Cartan's commutation coefficients \cite{cartan1922,misner73} (which are given, if we consider a vector field to the generator of a flow such that $\e\alpha=\sfrac{\partial\;}{\partial z^\alpha}$, by the Lie bracket $C_{\beta\alpha}^\delta\e\delta=L\sq{\e\beta,\e\alpha}=\sfrac{\partial\;}{\partial z^\alpha}\e\beta-\sfrac{\partial\;}{\partial z^\beta}\e\alpha$, which in our formalism reads $C_{\beta\alpha}^\delta\e\delta=(\con_{\alpha\beta}^\delta-\con_{\beta\alpha}^\delta)\e\delta$). 

Even without such generalization this is quite compelling. The electromagnetic field, viewed as an imaginary perturbation to the standard spacetime metric, appears as an imaginary term --- just the dual of the electromagnetic tensor $F_{\alpha\beta}=\partial_\alpha A_\beta-\partial_\beta A_\alpha$ --- in the Christoffel symbol $\con$. 

This is all very well conceptually, but what do such imaginary terms mean physically? Let us look at the motion and physical forces they create. 

To do this we ask what geodesic paths are followed by a particle in such a field. We first need the 4-momentum of a test particle. Continuing the idea above, where electromagnetism constitutes an imaginary perturbation of the metric, let us propose an electronic charge $e$ to constitute an imaginary perturbation of the classical 4-momentum, as
\begin{align}\label{imu}
u^\alpha=\gamma\cc{(m+\im e\rho)c,m\v v}\;,
\end{align}
where $\gamma$ is the usual Lorentz factor, and $\rho$ is a small quantity with the units of $mass/charge$. 
For instance this constant might be $\rho=1/\sqrt{\eps_0{\sf G}}$ where $\eps_0$ is the vaccum permittivity and $\sf G$ is the gravitational constant, which would give $e\rho\sim6.6\times10^{-9}kg$ in SI units (with $e$ being the elementary unit of charge). 

Since we have both real and imaginary components we must ask how we define a geodesic. Say we have a function $\phi(z)$ considered along a path $z(s)$, and consider choosing a path such that the imaginary part of $\phi$ is fixed, that is
\begin{align}\label{steep}
\imp{\left.\sfrac{d\;}{ds}\phi(z(s))\right|_{s=0}}=0\;.
\end{align}
This is important when integrating along a function containing a term like $e^{\phi}$, and seeking a non-oscillating, steepest descent, path for integration --- a `world line' for our test particle as a steepest descent solution of some appropriate integral; we remark again on this at the end of this letter. For now, consider a couple of simple examples. 

\begin{example}
If $\phi=\hf i (u-i z)^2=\hf i (u^2-z^2)+z\cdot u$, then \cref{steep} implies
\begin{align}\label{eikham}
0&=\imp{\left.\sfrac{d\;}{ds}\phi(z(s))\right|_{s=0}}\nonumber\\
&=\hf\sfrac{d\;}{ds}(u^2-z^2)
=u\cdot\dot u-z\cdot\dot z\;,
\end{align}
denoting the (local) derivative with respect to $s$ with a dot. The obvious solution of this is $\dot z=u$, $\dot u=-z$, defining a Hamiltonian system on the space-time vector $z=ct\e0+x^i\e i$ and 4-momentum vector $u=\sfrac Ec\e0+p^i\e i$. 
\end{example}

\begin{example}
Now consider if $\phi(z(s))=i u$, and $u=\dot z$. Then \cref{steep} implies
\begin{align}\label{geodesiC}
0&=\imp{\left.\sfrac{d\;}{ds}\phi(z(s))\right|_{s=0}}\nonumber\\
&=\rep{\dot u}=\rep{\dot z^\beta\sfrac{d\;}{dz^\beta}u}=\rep{u^\beta u^\alpha_{;\beta}}\e\alpha\;.
\end{align}
\end{example}

If $u$ is real then the steepest descent condition \cref{geodesiC} is just the familiar parallel transport (or geodesic) equation $u^\beta u^\alpha_{;\beta}=0$. In our case, by \cref{imu}, $u$ is complex. 

Say a particle follows a path $x(\tau)$ with tangent vector $x'(\tau)=u$, and curvature $x''(\tau)=x'(\tau)\cdot\sfrac{d\;}{d x}x'(\tau)=u^\beta u_{;\beta}\e\alpha$. Following \cref{geodesiC} let us define parallel transport as requiring only the vanishing of the real part of this. This means there is no displacement of $u$ away from the path, while the wandering imaginary part permits rotation of $u$ about the path (reminiscent of Weyl's early attempts to generalize Einstein's theory). We therefore require the real part of
\begin{align}\label{geocom}
u^\beta u^\alpha_{;\beta}&=u^\beta u^\alpha_{,\beta}+u^\beta u^\lambda\con^\alpha_{\lambda\beta}\;,
\end{align}
to vanish. A lengthy but straightforward calculation (see \cite{rasp2020long}) yields
\begin{align}
-v^i_{,0}
&\approx\hf c\mu_gG_i -\sfrac {ec\rho}m\mu_e (\id^{ij}E_j + v^j \eta^{ik}_j B_k)\;,
\end{align}
up to correction terms of order $\mu_g e^2\rho^2$. In a spatial geometry the magnetic term is the familiar cross-product $v^j \eta^{ik}_j B_k=(v\times B)_i$. Taking the gravitational component only we have the Newtonian force
\begin{align}\label{eomg}
cmv^i_{,0}\approx-mG_i +\dots\;,\qquad\qquad\qquad
\end{align}
up to a higher order perturbation `$+\ord{e^2\rho^2}$' from the electromagnetic contribution to the particle's 4-momentum, with the `0' subscript denoting the derivative $v_{,0}=\sfrac1c\sfrac{\partial v}{\partial t}$.
Taking only the electromagnetic component, and letting $\rho\mu_e=1/c^2$, provides the Lorentz force
\begin{align}\label{eome}
cmv^i_{,0}\approx {e} (E_i + v^j \eta^{ik}_j B_k)+\dots\;,\qquad
\end{align}
for a test particle with mass $m$, charge $e$, and velocity $v$. 
So our imaginary perturbations of the metric, yielding imaginary Christoffel symbols components, result in real forces corresponding to the familiar and electromagnetic forces, alongside the gravitational forces from real perturbations.

\bigskip

We jumped straight into possible applications to physical law above, but now let us return to the mathematics itself. 
The importance of the mixture is in providing an algebraic product with which we can pick apart expressions like the differential of a function with respect to  a variable $z$,
\begin{align}\label{df1}
df= dz\sfrac{d\;}{dz}f+\ord{dz^2}\;,
\end{align}
in less conventional ways than have become standard. 
It allows us to ask, for instance, when the expression \cref{df1} can be formed such that the function $f$, variable $z$, and derivative $\sfrac{d\;}{dz}f$, are all of the same type (i.e. belong to the same `space'). 
In one dimension the relation \cref{df1} is fundamental to differential and integral calculus, leading to a rigorous notion of a derivative, implicit differentiation and the `chain rule'. 

In more than one dimension this is non-trivial, requiring either specially defined products or special conditions relating $f$ and $z$. In two dimensions the relation \cref{df1} holds as written if $z$ and $f$ are complex variables, but only if $f$ satisfies the Cauchy-Riemann conditions. In higher dimensions the form \cref{df1} becomes too restrictive to hold for all but trivial (i.e. constant) functions, {\it unless} we let $\sfrac{d\;}{dz}f$ consist not only of the obvious derivative operator acting on $f$, namely $\sfrac{\partial\;}{\partial z}f$, but also an error term we call $\con(f)$, such that \cref{df1} can be written as
\begin{align}\label{df2}
df=dz\big(\sfrac{\partial\;}{\partial z}f+\con(f)\big)+\ord{dz^2}\;.
\end{align}
Let's say we can associate the error term $\con(f)$ with the derivative of the basis in which $f$ and $z$ are expressed. That is, to obtain \cref{df1} in its augmented form \cref{df2}, we will permit variation of the coordinate basis, and in doing so make the requirement of differentiability a {\it source term} for curvature of the underlying space. We refer to this as {\it analytic curvature}. 


Now, in complex variables there is one `conjugate' that carries out a symmetry operation on the basis, differentiability amounts to vanishing of the conjugate derivative. In higher dimensions we accrue more of these 'conjugacy' or {\it symmetry} operations, but importantly, analytic curvature does not require all of them to vanish, only enough that a curvature term $\Gamma$ can absorb any remaining errancy.  Let us identify an {\it adjoint} operation on $z$, labelled $\adj z$, that must vanish, that is
\begin{align}
0
&=d\adj z\sfrac{d\;}{d\adj z}f\nonumber\\
&=\e{\adj\gamma} dz^{\adj\gamma}\ec{\adj\beta}\e\alpha f^\alpha_{;\adj\beta}\nonumber\\
&=\e{\adj\gamma}\delta z^{\adj\gamma}\e\mu\eta^{\adj\beta\mu}_\alpha\bb{\sfrac{\partial\;}{\partial z^{\adj\beta}}f^\alpha+\con^\alpha_{\lambda\adj\beta}f^\lambda}
\end{align}
which holds for any $d\adj z=\e{\adj\gamma} dz^{\adj\gamma}$ if $
0
=\e\mu\eta^{\adj\beta\mu}_\alpha f^\alpha_{;\adj\beta}\nonumber
$, and since this must vanish for each $\mu$-indexed component we have
\begin{align}\label{anacond}
0=\eta^{\adj\beta\mu}_\alpha f^\alpha_{;\adj\beta}
\;.
\end{align}
We call this the {\it analyticity condition}. 
In essence \cref{anacond} is the extension of the Cauchy-Riemann equations to our covariant geometries, and indeed it reduces to them for commutative complex functions. 

For an example of how this restricts physical laws, above we defined a potential $h$ and saw how its derivatives related to electromagnetic terms. If we define the vector field $f=\frac{dh}{dz}=\alpha-\v e+i\v b$, then analyticity of $f$ according to \cref{anacond} gives precisely the (macroscopic) Maxwell equations for $\v e$ and $\v b$ with a source term $\frac{d\alpha}{dz}$. 

Writing this more generally, it is convenient to let 
 $\eta_{\lambda}^{\gamma\beta} \con_{\alpha\beta}^\lambda=\n H^\gamma_\alpha$ for some scalar $\n$ and some normalized tensor $ H^\gamma_\alpha$, then we can express analyticity of $f$ as requiring
\begin{align}
0= \sfrac{\partial\;}{\partial\adj z}f=\eta^{\gamma\beta}_\alpha f^\alpha_{;\adj\beta}
&= (\eta_{\alpha}^{\gamma\adj\beta} \partial_{,\beta}+\n\hat H^\gamma_\alpha)f^\alpha\;.
\end{align}
This looks superficially like Dirac's equation for an electron of mass $\m$ if $\n=\m c/\im\hbar$, but to confirm this beyond superficiality requires something stronger, namely that $f$ satisfies the Klein-Gordon equations. Those equations correspond to a strong form of analyticity conditions for $\sfrac{\partial\;}{\partial z}f$, in which only terms like $\partial_t^2$ and $\delta_x^2$ appear (i.e. no mixed $\delta_{tx}$ terms), and this requires rather strong conditions on the bases which translate into conditions on the mixture of
\begin{align}
\eta^{\gamma0}_\alpha=1^\gamma_\alpha\qquad\&\qquad \hat H=-H
\end{align}
where $1^\gamma_\alpha$ is the identity matrix. Then the matrices $\eta^{\gamma1}_\alpha$, $\eta^{\gamma2}_\alpha$, $\eta^{\gamma3}_\alpha$, $H^\gamma_\alpha$, behave algebraically as the Dirac matrices, that is we obtain, letting $\tens\eta^\beta$ and $\tens H$ be the components of $4\times4$ matrices $\eta^{\gamma\beta}_\alpha$ and $H^\gamma_\alpha$, that the matrices are roots of the identity
\begin{subequations}
\begin{align}\label{dirmat1}
\tens1&=\tens\eta^0\tens\eta^0=\tens\eta^i\tens\eta^i=
\tens H\hspace{0.02cm}{\tens H}\;,\qquad\qquad\;\;
\end{align}
and moreover commute with $\tens{\eta^0}$, 
\begin{align}\label{dirmat2}
\tens0&=\tens\eta^0\tens\eta^i-\tens\eta^i\tens\eta^0=
\tens H\hspace{0.02cm}\tens\eta^0-\tens\eta^0{\tens H}\;,\quad\;\;
\end{align}
and otherwise anti-commute
\begin{align}\label{dirmat3}
\tens0&=\left.\tens\eta^i\tens\eta^j+\tens\eta^j\tens\eta^i\right|_{i\neq j}=
\tens\eta^i{\tens H}+\tens H\hspace{0.02cm}\tens\eta^i\;.
\end{align}
\end{subequations}
Thus we have the Dirac equations \cite{dirac28,diracqm}, with the mixture $\eta$ and curvature term $H=\eta\con$ assuming the role of the Dirac matrices.

Of course one may find various elegant expressions of such laws in terms of geometric algebras or other formalisms. 
We have derived these laws from a differentiability condition, but many different expressions of the Dirac equations have been found over the years in different geometric or tensor algebras. The compelling feature of these investigations is how easily the familiar forms of these laws arise in an algebra and calculus based around the {mixture} $\eta$. 
Moreover the bases of Dirac's equations are revealed to be {\it not} spacial bases themselves, but products thereof in a suitable algebra, and yet still one that does not correspond to familiar geometrical space-time. 
The various possible symmetries of the mixture $\eta$ of bases (rather than the bases themselves) now take centre stage in determining the forms of differential laws.

Here, the familiar quantum wave expressions arise from analyticity conditions, while gravitational or electromagnetic expressions come from curvature (i.e. path integral and geodesic) conditions. The role of the `time' $t$ in the quantum and gravitational expressions is seen here to be not fundamentally different, rather it is where in the analysis (or calculus) the differential relations originate that changes the perceived meaning.  

The details of all these calculations and deeper investigation are to be found in \cite{rasp2020long}. The real intent of that study is to explore the possibility of extending phase integral methods in general to non-scalar variables, in particular so that the integral $\int dz\;g$, of some function $g$ with an antiderivative $f$, can be made to satisfy
\begin{align}\label{fund}
\int_a^b\!\!dz\;g=\int_a^b\!\!dz\;\sfrac{d\;}{dz}f=\int_{f(a)}^{f(b)}\!\!\!df=f(b)-f(a)\;
\end{align}
and thus be path independent. 
Such a form of integrability would permit the higher dimensional application of integral methods learned from the complex plane, from residues to steepest descents and stationary phases, to variational concepts from Feynman's path integrals to wave asymptotics or optimization problems. 

The equation \cref{dfcon} expresses a derivative function $g=\sfrac{df}{dz}$ using the mixture tensor, with which the equation \cref{anacond} is able to derive analyticity conditions. In between lie a whole raft of physical law based just on the algebra needed to connect these two links. 
If we have a function $g=e^{\phi(z)}$ considered along a path $z=z(s)$ for $s\in\mathbb R$, then to leading order in $s$, this path is non-oscillating (in stationary phase) at a point $s=0$ where \cref{steep} holds. 

What is suggested by these calculations is that the fundamental fields and constant of physics are not fundamental to the universe as a distinct physical entity, but are made inevitable by the language we use to describe it. They become fundamental once we insist on a mathematics that uses differentiable or integrable quantities. The resulting algebra and calculus are then so confined that things like Maxwell's, Dirac's, and Schr\"odinger's, laws are a matter of mathematical necessity. The patterns we see in the world (these fundamental laws) are set largely by the spectrum (of differentiable functions) we choose to view it in. 

We have not yet gone so far as to restrict the the parameters appearing in these laws (masses, charges, or coupling constants, for example), or even determine whether these are merely geometric or are specific to the universe in a given state at a given time. One might hope that such further inferences might be possible. 

Despite rather cursively visiting so many physical laws here, I am not trying to claim any solution to the puzzle of physical laws, but merely remove one element from the puzzle. I claim that many of the fundamental laws are due simply to seeking differentiable quantities to describe the world, that many of its symmetries (and proposed higher dimensions) can be described by the appropriate {\it mixture} of non-trivial space-time geometries. In doing so I hope merely to help separate those mathematical equations and quantities that are intrinsic to the physics of our particular universe, from those inherited from the mathematics we have derived to describe it. 


\bibliography{../grazcat}

\begin{thebibliography}{1}

\bibitem{cartan1922}
E.~Cartan.
\newblock Sur une g\'en\'eralisation de la notion de courbure de {R}iemann et
  les espaces \`a torsion.
\newblock {\em C. R. Acad. Sci.}, 174:593--595, 1922.

\bibitem{cg07}
W.~N. Cottingham and D.~A. Greenwood.
\newblock {\em An introduction to the standard model of particle physics}.
\newblock Cambridge University Press, 2007.

\bibitem{dirac28}
P.~A.~M. Dirac.
\newblock The quantum theory of the electron.
\newblock {\em Proc. R. Soc. A}, 1928.

\bibitem{diracqm}
P.~A.~M. Dirac.
\newblock {\em The Principles of Quantum Mechanics}.
\newblock Oxford University Press, 1930.

\bibitem{rasp2020long}
M.~R. Jeffrey.
\newblock Complex geometry and fundamental physical law.
\newblock {\em arXiv:1910.04264}, 2019.

\bibitem{misner73}
C.~W. Misner, K.~S. Thorne, and J.~A. Wheeler.
\newblock {\em Gravitation}.
\newblock San Francisco: W. H. Freeman, 1973.

\bibitem{yangmills}
C.~N. Yang and R.~K. Mills.
\newblock Conservation of isotopic spin and isotopic gauge invariance~.
\newblock {\em Physical Review}, 96(1):191--5, 1954.

\end{thebibliography}
\bibliographystyle{plain} 

\clearpage\clearpage

\end{document}